\documentstyle[12pt]{article}

\input{epsf.sty}
\def\axowidth{0.5 }
\def\axoscale{1.0 }
\def\axoxoff{0 }
\def\axoyoff{0 }

\def\Gluon(#1,#2)(#3,#4)#5#6{
\put(\axoxoff,\axoyoff){
}
}

\def\Photon(#1,#2)(#3,#4)#5#6{
\put(\axoxoff,\axoyoff){
}
}

\def\ZigZag(#1,#2)(#3,#4)#5#6{
\put(\axoxoff,\axoyoff){
}
}

\def\PhotonArc(#1,#2)(#3,#4,#5)#6#7{
\put(\axoxoff,\axoyoff){
}
}

\def\GlueArc(#1,#2)(#3,#4,#5)#6#7{
\put(\axoxoff,\axoyoff){
}
}

\def\ArrowArc(#1,#2)(#3,#4,#5){
\put(\axoxoff,\axoyoff){
}
}

\def\LongArrowArc(#1,#2)(#3,#4,#5){
\put(\axoxoff,\axoyoff){
}
}

\def\DashArrowArc(#1,#2)(#3,#4,#5)#6{
\put(\axoxoff,\axoyoff){
}
}

\def\ArrowArcn(#1,#2)(#3,#4,#5){
\put(\axoxoff,\axoyoff){
}
}

\def\LongArrowArcn(#1,#2)(#3,#4,#5){
\put(\axoxoff,\axoyoff){
}
}

\def\DashArrowArcn(#1,#2)(#3,#4,#5)#6{
\put(\axoxoff,\axoyoff){
}
}

\def\ArrowLine(#1,#2)(#3,#4){
\put(\axoxoff,\axoyoff){
}
}

\def\LongArrow(#1,#2)(#3,#4){
\put(\axoxoff,\axoyoff){
}
}

\def\DashArrowLine(#1,#2)(#3,#4)#5{
\put(\axoxoff,\axoyoff){
}
}

\def\Line(#1,#2)(#3,#4){
\put(\axoxoff,\axoyoff){
}
}

\def\DashLine(#1,#2)(#3,#4)#5{
\put(\axoxoff,\axoyoff){
}
}

\def\CArc(#1,#2)(#3,#4,#5){
\put(\axoxoff,\axoyoff){
}
}

\def\DashCArc(#1,#2)(#3,#4,#5)#6{
\put(\axoxoff,\axoyoff){
}
}

\def\Vertex(#1,#2)#3{
\put(\axoxoff,\axoyoff){
}
}

\def\Text(#1,#2)[#3]#4{
\dimen0=\axoxoff \unitlength
\dimen1=\axoyoff \unitlength
\advance\dimen0 by #1 \unitlength
\advance\dimen1 by #2 \unitlength
\makeatletter
\@killglue\raise\dimen1\hbox to\z@{\kern\dimen0 \makebox(0,0)[#3]{#4}\hss}
\ignorespaces
\makeatother
}

\def\BCirc(#1,#2)#3{
\put(\axoxoff,\axoyoff){
}
}

\def\GCirc(#1,#2)#3#4{
\put(\axoxoff,\axoyoff){
}
}

\def\EBox(#1,#2)(#3,#4){
\put(\axoxoff,\axoyoff){
}
}

\def\BBox(#1,#2)(#3,#4){
\put(\axoxoff,\axoyoff){
}
}

\def\GBox(#1,#2)(#3,#4)#5{
\put(\axoxoff,\axoyoff){
}
}

\def\Boxc(#1,#2)(#3,#4){
\put(\axoxoff,\axoyoff){
}
}

\def\BBoxc(#1,#2)(#3,#4){
\put(\axoxoff,\axoyoff){
}
}

\def\GBoxc(#1,#2)(#3,#4)#5{
\put(\axoxoff,\axoyoff){
}
}

\def\SetOffset(#1,#2){\def\axoxoff{#1 } \def\axoyoff{#2 }}

\def\fsize{10 }

\def\PText(#1,#2)(#3)[#4]#5{
\ifx#4 lt{\def\fmode{0 }}\else{
\ifx#4 tl{\def\fmode{0 }}\else{
\ifx#4 lb{\def\fmode{2 }}\else{
\ifx#4 bl{\def\fmode{2 }}\else{
\ifx#4 l{\def\fmode{1 }}\else{
\ifx#4 rt{\def\fmode{6 }}\else{
\ifx#4 tr{\def\fmode{6 }}\else{
\ifx#4 rb{\def\fmode{8 }}\else{
\ifx#4 br{\def\fmode{8 }}\else{
\ifx#4 r{\def\fmode{7 }}\else{
\ifx#4 t{\def\fmode{3 }}\else{
\ifx#4 b{\def\fmode{5 }}\else{ \def\fmode{4 } }\fi
}\fi}\fi}\fi}\fi}\fi}\fi}\fi}\fi}\fi}\fi}\fi
\put(#1,#2){\makebox(0,0)[]{\special{"/pfont findfont /fsize scale setfont
\axoxoff \axoyoff #3 \fmode \fsize (#5) ptext }}}  }

\def\GOval(#1,#2)(#3,#4)(#5)#6{
\put(\axoxoff,\axoyoff){
}
}

\def\Oval(#1,#2)(#3,#4)(#5){
\put(\axoxoff,\axoyoff){
}
}

\let\eind=]

\def\kromme(#1,#2)#3{#1 #2 \ifx #3\eind\else\expandafter\kromme\fi#3}

\def\LogAxis(#1,#2)(#3,#4)(#5,#6,#7,#8){
\put(\axoxoff,\axoyoff){
}
}

\def\LinAxis(#1,#2)(#3,#4)(#5,#6,#7,#8,#9){
\put(\axoxoff,\axoyoff){
}
}

\title{Axion Decay in a Constant Electromagnetic Background Field and at Finite Temperature using World-line Methods}
\renewcommand{\thefootnote}{\fnsymbol{footnote}}
\author{Michael Haack\footnote{michael@hera2.physik.uni-halle.de}$\hspace{.15cm}^{,}\hspace{.025cm}^{a}$ \and Michael G. Schmidt\footnote{M.G.Schmidt@thphys.uni-heidelberg.de}$\hspace{.15cm}^{,}\hspace{.025cm}^{b}$}
\date{{\small $^{a}$ {\em Institut f\"ur Theoretische Physik, Martin-Luther-Universit\"at Halle-Wittenberg, \vspace{.3cm} Selkestr.~9, D-06099 Halle, Germany} \linebreak $^{b}${\em Institut f\"ur Theoretische Physik, Universit\"at Heidelberg, Philosophenweg 16, D-69120 Heidelberg, Germany}}}
\begin{document}
\maketitle

\begin{abstract}
We investigate the radiative decay of the axion into two photons in an external electromagnetic field to one loop order. Our approach is based on the world-line formalism, which is very suitable to take into account the external field to all orders. Afterwards we discuss how the calculation could be generalized to finite temperature. 
\end{abstract} 

\renewcommand{\thefootnote}{\arabic{footnote}}
\setcounter{footnote}{0}

\section{Introduction}
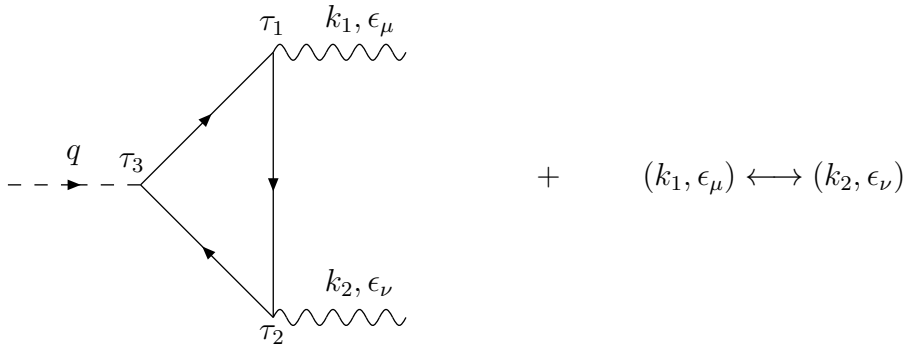
\begin{figure}[h]
\begin{center}
\begin{picture}(350,150)(0,0)
  \DashArrowLine(10,75)(60,75)5
  \put(32,85){$q$}
  \put(52,82){$\tau_{3}$}
  \ArrowLine(60,75)(110,125)
  \ArrowLine(110,25)(60,75)
  \ArrowLine(110,125)(110,25)
  \Photon(110,125)(160,125)35
  \put(105,134){$\tau_{1}$}
  \Photon(110,25)(160,25)35
  \put(105,17){$\tau_{2}$}
  \put(130,135){$k_{1},\epsilon_{\mu}$}
  \put(130,35){$k_{2},\epsilon_{\nu}$}
  \put(210,75){$+$}
  \put(250,75){$(k_{1},\epsilon_{\mu}) \longleftrightarrow (k_{2},\epsilon_{\nu})$}
\end{picture}
\caption{Triangle graph}
\label{fig1}
\end{center}
\end{figure}

In order to solve the strong CP problem Peccei and Quinn [1] proposed that the full lagrangian of the extended standard model is invariant under a global chiral $U(1)$ symmetry, which however is spontaneously broken, leading to a pseudoscalar Goldstone boson [2], the axion. This particle couples to fermions, depending on the model either to the known fermions of the standard model (DFSZ model [3]) or to some exotic heavy new fermions (KSVZ model [4]). In the latter case the axions couple to ordinary matter and radiation mainly by means of the triangle graph (fig. \ref{fig1}), leading to an effective two-photon (or two-gluon) coupling.

Although the original Peccei-Quinn model, which assumed a symmetry breaking at the electroweak scale, is 'ruled out' by experiment [5], models with a much larger breaking scale, leading to light and weakly coupled, so called invisible, axions are still of great interest. Their astrophysical impact, especially their influence on stellar evolution, is discussed e.g. in [6-8]. 

In the present paper we study the axion-two-photon coupling in the presence of a constant homogeneous electromagnetic background field. Thus we assume a pseudoscalar coupling of the axion field $\phi'$ to fermions, $\lambda' \phi' \overline{\psi} \gamma_{5} \psi$, although the Goldstone character of the axion would in general require a derivative coupling. It is argued, however, in [6] and references therein, that the pseudoscalar coupling leads to the correct results as long as only one axion is attached to a fermion line. This is obviously the case for the triangle graph.  

Our calculation is based on the world-line formalism [9-17]. This is a first quantized formalism, originally used for one loop calculations in quantum field theory. The generalisation to higher loop orders is however possible as elaborated in the scalar theory and particularly in QED (see ref. [14-16]). Although the idea of dealing with particles on a world-line is quite old [9], it has experienced a renaissance through the work of Bern and Kosower [10], who derived a set of new Feynman rules for one loop gluon scattering amplitudes from string theory, which thus inherited its superior organisation of perturbative calculations. Strasslers observation that the new set of rules can alternatively be derived through pathintegral methods has been very useful [11]. In this approach one writes the one loop effective action as a (super)particle path integral and evaluates this path integral using world-line Green functions appropriate to a one dimensional field theory on the circle. 

The world-line method is particularly useful in situations with a constant background field [16] and has been successfully applied to the calculation of photon splitting in a strong magnetic field [18].  The description of scalar and pseudoscalar couplings on the world-line is given in [19-21]. In [20] the world-line lagrangian was derived by analogy to the  expression in the second order description of the fermionic one loop effective action in a background of vector, axial vector, scalar and pseudoscalar fields. Thus it was mathematically necessary to introduce two further bosonic and fermionic degrees of freedom. This procedure was clarified in [21], where the world-line action for a fermion coupled to vector, scalar and pseudoscalar fields is derived from an eight dimensional Dirac operator in six spacetime dimensions coupled to a gauge field whose first four components are those of the vector gauge field while the fifth and sixth components are the scalar respectively pseudoscalar field. The Dirac fermion however only depends on the first four coordinates. This interpretation is possible if one doubles the fermion degrees (without altering the value of the effective action) and introduces the six Gamma matrices

\begin{equation}
\Gamma_{\mu} = \left( \begin{array}{cc}
                      0 & \gamma_{\mu} \\
                      \gamma_{\mu} & 0 
                      \end{array}
               \right)
\hspace{.2cm} , \hspace{.2cm}
\Gamma_{5} = \left( \begin{array}{cc}
                    0 & i I \\
                    -iI & 0 
                    \end{array}
             \right)
\hspace{.2cm} , \hspace{.2cm}
\Gamma_{6} = \left( \begin{array}{cc}
                    0 & \gamma_{5} \\
                    \gamma_{5} & 0
                    \end{array}
             \right)   \label{eq0}
\end{equation}
which obey the usual anticommutation relations 

\begin{equation}
\{\Gamma_{A}, \Gamma_{B} \} = 2 \delta_{AB} I_{8 \times 8}   \label{eq0a}
\end{equation}
The above mentioned six spacetime dimensional Diracoperator then has the form:

\begin{equation}
\Sigma = \Gamma_{\mu} (p_{\mu} - e A_{\mu}) - \Gamma_{5} \lambda \phi - \Gamma_{6} \lambda' \phi'  \label{eq0b}
\end{equation}
Due to the presence of the pseudoscalar field, the effective action has both a real and an imaginary part, depending on whether an even or an odd number of pseudoscalar particles are involved. The real part of the effective action can be calculated as

\begin{equation}
\Gamma_{\Re} = -\frac{1}{2} \ln \mbox{Det} [\Sigma] = - \int_{0}^{\infty} \frac{ds}{s} \int_{P} {\cal{D}}x \int_{A} {\cal{D}} \psi \, e^{-S}  \label{eq0c}
\end{equation}
where $P$ and $A$ denote periodic and antiperiodic boundary conditions resp. and $S$ is the world-line action, which for the case at hand (i.e. in the formulas of [20] we discard the axial vector contribution, introduce the fermion mass by setting the scalar field to $\phi \rightarrow m/\lambda$ and integrate out the auxiliary variables $x_{5}$ and $x_{6}$ via their 'equation of motion') is

\begin{eqnarray}
S & = & \int_{0}^{s} d\tau \, \{ \frac{\dot{x}^{2}}{4} + \frac{1}{2} \psi \cdot  \dot{\psi} + \frac{1}{2} \psi_{5} \dot{\psi_{5}} + \frac{1}{2} \psi_{6} \dot{\psi_{6}} + m^{2} + \lambda'^{2} \phi'^{2} - 2 i \lambda' \psi_{6} \psi_{\mu} \partial_{\mu} \phi'  \nonumber \\
& &  \hspace{1.2cm} + i e (\dot{x}_{\mu} A_{\mu} + 2 \psi_{\mu} \psi_{\nu} \partial_{\nu} A_{\mu}) \} \label{eq1}
\end{eqnarray}
As the triangle graph involves only one axion field it contributes to the imaginary part of the effective action. According to [20] its derivative with respect to the pseudoscalar field is given by\footnote{The normalization stems from a calculation of the anomaly in four dimensions (see e.g. [32], p.552) and agrees with the one of [21].}    

\begin {equation}
\frac{\delta}{\delta \phi'} i \Gamma_{\Im} = - \int_{0}^{\infty} \frac{ds}{s} \int_{P} {\cal{D}}x \int_{P} {\cal{D}} \psi \, \Omega_{\phi'} e^{-S}   \label{eq1a}
\end{equation}
where $S$ is again given by (\ref{eq1}), the boundary conditions for the fermions have changed to periodic ones (as a consequence of the appearance of $\gamma_{5}$ in the usual Feynman calculation) and the insertion operator is (again for the case at hand)

\begin{equation}
\Omega_{\phi'} = - i \lambda' \int_{0}^{s} d\tau \, \left( \frac{1}{2} \psi_{\mu} \dot{x}_{\mu} \psi_{6} - i m \psi_{5} \psi_{6} \right)   \label{eq2}
\end{equation}
When solving the fermionic pathintegral by means of Wick-contractions with the world-line Green functions, we have to take into account the periodic boundary conditions of the fermions. There are now fermionic zero modes and the world-line supersymmetry is no longer broken by the boundary conditions. Thus we have

\begin{equation}
G_{F}^{periodic} = \dot{G}_{B}  \hspace{.5cm} , \hspace{.5cm} G_{B}(\tau_{1}, \tau_{2}) = |\tau_{1} - \tau_{2}| - \frac{(\tau_{1} - \tau_{2})^{2}}{s}   \label{eq3}
\end{equation}
where $G_{F/B}$ means the fermionic respectively bosonic Green function on the world-line.

\section{The case \{$F_{\mu \nu} \neq 0, \hspace{.3cm} T, \mu = 0$\}}

The amplitude for the coupling of the axion to the electromagnetic field is given in the world-line formalism by:

\begin{eqnarray}
{\cal M} & = & - \int_{0}^{\infty} \frac{ds}{s} \int_{P} {\cal D}y {\cal D} \tilde{\psi}_{\mu} {\cal D} \tilde{\psi}_{5} {\cal D} \tilde{\psi}_{6} \int d\psi^{(0)}_{\mu} d\psi_{5}^{(0)} d\psi_{6}^{(0)} \int dx^{(0)} \nonumber \\ 
& & \hspace{-1cm} (-i \lambda') \int_{0}^{s} d\tau' \, \left( \frac{1}{2} \psi_{\mu} \dot{y}_{\mu} \psi_{6} - i m \psi_{5} \psi_{6} \right) e^{i q y(\tau')} e^{i q x^{(0)}} e^{-m^2 s}     \label{eq4} \\
& & \hspace{-1cm} \exp \left\{-\int_{0}^{s} d\tau \, \left[ \frac{\dot{y}^{2}}{4} + \frac{1}{2} \psi \cdot \dot{\psi} + \frac{1}{2} \psi_{5} \dot{\psi_{5}} + \frac{1}{2} \psi_{6} \dot{\psi_{6}} + i e (\dot{y}_{\mu} A_{\mu} + 2 \psi_{\mu} \psi_{\nu} \partial_{\nu} A_{\mu}) \right] \right\}  \nonumber 
\end{eqnarray} 
where the integration over the zeromodes has been seperated ($x=x^{(0)} + y$, and $\int_{0}^{s} d\tau y^{\mu}(\tau) = 0$). One has to use $\psi_{\mu,5,6} = \tilde{\psi}_{\mu,5,6} + \psi_{\mu,5,6}^{(0)}$ in the integrand. As the integration over the fermionic zeromodes is only nonvanishing if the integrand contains a factor $\psi_{\mu}^{(0)} \psi_{\nu}^{(0)} \psi_{\alpha}^{(0)} \psi_{\beta}^{(0)} \psi_{5}^{(0)} \psi_{6}^{(0)}$, where the greek indices run from $1$ to $4$, producing an $\epsilon_{\mu \nu \alpha \beta}$, one realizes, that the first term of the inserted operator does in fact not contribute. One also notices, that $\int_{0}^{s} d\tau \, \psi_{\mu,5,6}^{(0)} \dot{\tilde{\psi}}_{\mu,5,6} = 0$ so that in the kinetic part of the fermionic world-line action the substitution $\psi_{\mu,5,6} \rightarrow \tilde{\psi}_{\mu,5,6}$ is allowed. 
    
As described in [16] we can seperate the electromagnetic field into the constant, homogeneous background field $\bar{F}$ and the outgoing photon fields $\tilde{F}$ (which are described like the axion by plane waves):
\begin{equation}
F_{\mu \nu} = \bar{F}_{\mu \nu} + \tilde{F}_{\mu \nu}    \label{eq5}
\end{equation}
Using the Fock-Schwinger-gauge for the constant field yields:
\begin{equation}
A_{\mu} (\tau) = \bar{A}_{\mu} + \tilde{A}_{\mu} (\tau) = \frac{1}{2} y_{\nu} \bar{F}_{\nu \mu} + \epsilon_{1 \mu} e^{i k_{1} x(\tau)} + \epsilon_{2 \mu} e^{i k_{2} x(\tau)}    \label{eq6} 
\end{equation}
We realize that the Fock-Schwinger-gauge is compatible with the Lorentz-gauge, so that we can assume $\epsilon_{i} \cdot k_{i} = 0 , \, i = 1,2$.
The influence of the constant background field can partially be incorporated by a redefinition of the Green functions and the fluctuation determinants. In doing so we have to keep in mind the periodic boundary conditions of the fermions (see appendix A). 

If we expand the interaction part of the exponential in the integrand of (\ref{eq4}) and keep only terms, which are linear in $\epsilon_{1}$ and $\epsilon_{2}$ and which do contribute after the integration of the fermionic zeromodes, we are left with the computation of the following expression:

\begin{eqnarray}
{\cal M} & = & - \int_{0}^{\infty} \frac{ds}{s} \int dx^{(0)} \int_{P} {\cal{D}}y {\cal{D}}\tilde{\psi}_{\mu} d \psi^{(0)}_{\mu}  d \psi^{(0)}_{5} d \psi^{(0)}_{6}  \nonumber \\
& & e^{-m^{2} s} (-i \lambda ') \int_{0}^{s} d\tau_{1} d\tau_{2} d\tau_{3} \, (-i \psi_{5}^{(0)} \psi_{6}^{(0)} m) e^{i q x^{(0)}} e^{i q y(\tau_{3})}  \nonumber \\
& & \frac{1}{2} \left[ i e \tilde{A}_{\mu}(\tau_{1}) \dot{y}_{\mu}(\tau_{1}) - i e (\psi_{\mu}^{(0)} + \tilde{\psi}_{\mu}(\tau_{1})) \tilde{F}_{\mu \nu} (\tau_{1}) (\psi_{\nu}^{(0)} + \tilde{\psi}_{\nu}(\tau_{1})) \right]   \nonumber \\
& & \left[ i e \tilde{A}_{\mu}(\tau_{2}) \dot{y}_{\mu}(\tau_{2}) - i e (\psi_{\mu}^{(0)} + \tilde{\psi}_{\mu}(\tau_{2})) \tilde{F}_{\mu \nu} (\tau_{2}) (\psi_{\nu}^{(0)} + \tilde{\psi}_{\nu}(\tau_{2})) \right]   \nonumber \\
& & \left(1 + i e s \psi_{\mu}^{(0)} \bar{F}_{\mu \nu} \psi_{\nu}^{(0)} - \frac{1}{2} e^{2} s^{2} (\psi_{\mu}^{(0)} \bar{F}_{\mu \nu} \psi_{\nu}^{(0)})^{2} \right)   \nonumber \\
& & \exp\left\{ -\int_{0}^{s} d\tau \, \left[ \frac{\dot{y}^{2}}{4} + \frac{1}{2} i e y_{\nu} \bar{F}_{\nu \mu} \dot{y}_{\mu} + \frac{1}{2} \tilde{\psi} \cdot \dot{\tilde{\psi}} - i e \tilde{\psi}_{\mu} \bar{F}_{\mu \nu} \tilde{\psi}_{\nu} \right] \right\}   \label{eq7}
\end{eqnarray}
where we have used $\int d \tau \, \psi_{\mu}^{(0)} \bar{F}_{\mu \nu} \tilde{\psi}_{\nu} (\tau) = 0$ and $\int {\cal{D}}\tilde{\psi}_{5/6} \exp\{-\int_{0}^{s} d \tau \, \frac{1}{2} \tilde{\psi}_{5/6} \dot{\tilde{\psi}}_{5/6} \}$ $ = 1$. The $\tau$s are integrated over the circle with circumference $s$ and one $\tau$ can be set to zero for convenience because of translational invariance on the world-line (in our calculations this will be $\tau_{3}$). We now perform the Wick-contractions with the bosonic Green function\footnote{In the following ${\cal{G}}$ always means the bosonic Green function ${\cal{G}}_{B}$ (\ref{eqa2}).} in a constant background field, i.e. we apply the following rules:

\begin{eqnarray}
<y_{\mu} (\tau_{1}) y_{\nu} (\tau_{2})> & = & -{\cal{G}}_{\mu \nu}(\tau_{1}, \tau_{2})     \label{eq8} \\
<\dot{y}_{\mu} (\tau_{1}) \dot{y}_{\nu}(\tau_{2})> & = & -\dot{{\cal{G}}}'_{\mu \nu} (\tau_{1}, \tau_{2})   \label{eq9}  \\
<\dot{y}_{\mu}(\tau_{1}) e^{i k y(\tau_{2})}> & = & -i \dot{{\cal{G}}}_{\mu \nu} (\tau_{1}, \tau_{2}) k_{\nu} e^{i k y(\tau_{2})}   \label{eq10}  \\
<e^{i k_{1} y(\tau_{1})} e^{i k_{2} y(\tau_{2})} e^{i k_{3} y(\tau_{3})}> & = & \exp\{\sum_{m<n}k_{m \mu} k_{n \nu} {\cal{G}}_{\mu \nu} (\tau_{m}, \tau_{n}) \}  \label{eq11}  \\
<\tilde{\psi}_{\mu} (\tau_{1}) \tilde{\psi}_{\nu} (\tau_{2})> & = & \frac{1}{2} \dot{{\cal{G}}}_{\mu \nu} (\tau_{1}, \tau_{2})   \label{eq12}
\end{eqnarray}
where the dot and prime mean differentiation with respect to the first respectively second argument. We get rid of the term $\dot{{\cal{G}}}'_{\mu \nu}(\tau_{1}, \tau_{2}) = \frac{1}{2} \left(\partial_{\tau_{1}} {\cal{G}}'_{\mu \nu} (\tau_{1}, \tau_{2}) \right.$ $\left. + \partial_{\tau_{2}} \dot{{\cal{G}}}_{\mu \nu} (\tau_{1}, \tau_{2}) \right)$ by partial integration, using the symmetry properties:

\begin{eqnarray}
\dot{{\cal{G}}}_{\mu \nu}(\tau_{1}, \tau_{2}) & = & -\dot{{\cal{G}}}_{\nu \mu} (\tau_{2}, \tau_{1})  \label{eq13} \\
\dot{{\cal{G}}}_{\mu \nu}(\tau_{1}, \tau_{2}) & = & -{\cal{G}}'_{\mu \nu}(\tau_{1}, \tau_{2})   \label{eq14}
\end{eqnarray}
and noticing that the boundary terms vanish because of the periodicity of $\dot{{\cal{G}}}_{\mu \nu} (0, \tau_{2}) = \dot{{\cal{G}}}_{\mu \nu} (s, \tau_{2})$. 
The unbroken supersymmetry on the world-line leads to extensive cancellations in the calculation. We finally get as our result (setting $k_{3} = q$ for convenience):

\begin{equation}
{\cal M} = {\cal M}_{0} + {\cal M}_{1} + {\cal M}_{2}  \label{eq15}
\end{equation}
with

\begin{eqnarray}
{\cal M}_{0} & = & 4 m \lambda' e^{2} \int_{0}^{\infty} \frac{ds}{s} \int_{0}^{s} d\tau_{1} d\tau_{2} d\tau_{3} e^{-m^{2} s} (2 \pi)^{4} \delta(k_{1} + k_{2} + k_{3}) (4 \pi s)^{-2}  \nonumber \\
& & \exp\{\frac{1}{2}\sum_{i \neq j} k_{i} \cdot {\cal{G}} (\tau_{i}, \tau_{j}) \cdot k_{j} \} \epsilon_{\mu_{1} \nu_{1} \mu_{2} \nu_{2}} k_{1 \mu_{1}} \epsilon_{1 \nu_{1}} k_{2 \mu_{2}} \epsilon_{2 \nu_{2}}   \label{eq16} \\
& &   \nonumber \\
{\cal M}_{1} & = & 2 i m \lambda' e^{3} \int_{0}^{\infty} ds \int_{0}^{s} d\tau_{1} d\tau_{2} d\tau_{3} e^{-m^{2} s} (2 \pi)^{4} \delta(k_{1} + k_{2} + k_{3}) (4 \pi s)^{-2}  \nonumber \\
& & \exp\{\frac{1}{2}\sum_{i \neq j} k_{i} \cdot {\cal{G}} (\tau_{i}, \tau_{j}) \cdot k_{j} \}   \nonumber \\
& & \left[ 2 (k_{1} \cdot \hat{F} \cdot \epsilon_{1}) \left\{ (\epsilon_{2} \cdot \dot{{\cal{G}}} (\tau_{2}, \tau_{1}) \cdot k_{1}) + (\epsilon_{2} \cdot \dot{{\cal{G}}} (\tau_{2}, \tau_{3}) \cdot k_{3}) \right\}  \right.  \nonumber \\
& & \mbox{} - (k_{1} \cdot \hat{F} \cdot k_{2}) (\epsilon_{1} \cdot \dot{{\cal{G}}}(\tau_{1}, \tau_{2}) \cdot \epsilon_{2}) - (\epsilon_{1} \cdot \hat{F} \cdot \epsilon_{2}) (k_{1} \cdot \dot{{\cal{G}}}(\tau_{1}, \tau_{2}) \cdot k_{2})    \nonumber \\
& & \left. \mbox{} + 2 (k_{1} \cdot \hat{F} \cdot \epsilon_{2}) (\epsilon_{1} \cdot \dot{{\cal{G}}} (\tau_{1}, \tau_{2}) \cdot k_{2}) \right]   \label{eq17} \\
& & \mbox{} + (\epsilon_{1}, k_{1}) \leftrightarrow (\epsilon_{2}, k_{2})  \nonumber  \\
& &  \nonumber \\
{\cal M}_{2} & = & - \frac{1}{2} m \lambda' e^{4} \int_{0}^{\infty} ds \, s \int_{0}^{s} d\tau_{1} d\tau_{2} d\tau_{3} e^{-m^{2} s} (2 \pi)^{4} \delta(k_{1} + k_{2} + k_{3}) (4 \pi s)^{-2}  \nonumber \\
& & \exp\{\frac{1}{2}\sum_{i \neq j} k_{i} \cdot {\cal{G}} (\tau_{i}, \tau_{j}) \cdot k_{j} \} \bar{F}_{\mu \nu} \hat{F}_{\mu \nu}  \nonumber \\
& & \left[ \frac{1}{2} (\epsilon_{1} \cdot \dot{{\cal{G}}}(\tau_{1}, \tau_{2}) \cdot \epsilon_{2}) \left\{ (k_{1} \cdot \dot{{\cal{G}}} (\tau_{1}, \tau_{3}) \cdot k_{3}) - (k_{2} \cdot \dot{{\cal{G}}} (\tau_{2}, \tau_{3}) \cdot k_{3}) \right\} \right.  \nonumber \\
& & \mbox{} + 2 (\epsilon_{1} \cdot \dot{{\cal{G}}} (\tau_{1}, \tau_{3}) \cdot k_{3}) (\epsilon_{2} \cdot \dot{{\cal{G}}} (\tau_{2}, \tau_{1}) \cdot k_{1})  \nonumber \\
& & \left. \mbox{} + (\epsilon_{1} \cdot \dot{{\cal{G}}} (\tau_{1}, \tau_{3}) \cdot k_{3}) (\epsilon_{2} \cdot \dot{{\cal{G}}} (\tau_{2}, \tau_{3}) \cdot k_{3}) \right]  \label{eq18} \\
& & \mbox{} + (\epsilon_{1}, k_{1}) \leftrightarrow (\epsilon_{2}, k_{2})  \nonumber
\end{eqnarray}
where we have introduced the notation $\hat{F}_{\mu \nu} = \frac{1}{2} \epsilon_{\mu \nu \alpha \beta} \bar{F}_{\alpha \beta}$.

Before going on to the finite temperature case we want to specify our formula to the case described in [22], i.e. we neglect the axion mass and assume the crossed field case $(\vec{E} \perp \vec{B}, E = B $ or $\bar{F}_{\mu \nu} \bar{F}_{\mu \nu} = \bar{F}_{\mu \nu} \hat{F}_{\mu \nu} = 0)$. In the limit of a vanishing axion mass we additionally have $q \parallel k_{1} \parallel k_{2}$ because of energy-momentum conservation, that means $k_{1} = - \Lambda q$, $k_{2} = - (1 - \Lambda) q$.  

We see immediately that under these circumstances only ${\cal M}_{1}$ contributes to the amplitude. Furthermore we easily verify, that $\bar{F}_{\mu \alpha} \bar{F}_{\alpha \beta} \bar{F}_{\beta \nu} = 0$, so that we can use the expansions (\ref{eqa4}) and (\ref{eqa5}) in ${\cal M}_{1}$. In doing so we arrive at the following formula (see appendix B for some details):

\begin{eqnarray}
{\cal M} & = &  4 i m \lambda' e^{3} \int_{0}^{\infty} ds \, \int_{0}^{s} d\tau_{1} d\tau_{2} d\tau_{3} \, e^{-m^{2} s} (2 \pi)^{4} \delta(k_{1} + k_{2} + k_{3}) (4 \pi s)^{-2} \frac{1}{1-\Lambda} \nonumber \\
& & \exp\{\frac{1}{2}\sum_{i \neq j} k_{i} \cdot {\cal{G}} (\tau_{i}, \tau_{j}) \cdot k_{j} \} (k_{1} \cdot \hat{F} \cdot \epsilon_{1}) (\epsilon_{2} \cdot [\dot{{\cal{G}}}(\tau_{2}, \tau_{1}) - \dot{{\cal{G}}} (\tau_{2}, \tau_{3})] \cdot k_{2})  \nonumber \\
& & \mbox{} + (\epsilon_{1}, k_{1}) \leftrightarrow (\epsilon_{2}, k_{2})  \label{eq19} \\
& = &  \frac{1}{4} m \lambda' e^{4} \pi^{-2} (2 \pi)^{4} \delta(k_{1} + k_{2} + q) \int_{0}^{\infty} ds \, s^{2} \int_{0}^{1} d\sigma_{1} \int_{0}^{1-\sigma_{1}} d\sigma_{2} \, e^{-m^{2} s}   \nonumber  \\
& & \left( \frac{1}{1 - \Lambda} (f_{1 \mu \nu} \hat{F}_{\mu \nu}) (f_{2 \mu \nu} \bar{F}_{\mu \nu}) (\sigma_{1} - 2 \sigma_{1} \sigma_{2} - \sigma_{1}^{2}) + (\epsilon_{1}, k_{1}) \leftrightarrow (\epsilon_{2}, k_{2}) \right)  \nonumber \\
& & \exp\{ \frac{s^{3}}{3} e^{2} (q \cdot \bar{F} \cdot \bar{F} \cdot q) \, [- (\sigma_{1} - \sigma_{1}^{2})^{2} \Lambda^{2} - (\sigma_{2} - \sigma_{2}^{2})^{2} (1 - \Lambda)^{2}   \nonumber \\
& &  2 \sigma_{1} \sigma_{2} (1 - 3 \sigma_{1} - 3 \sigma_{2} + 2 \sigma_{1}^{2} + 2 \sigma_{2}^{2} + 3 \sigma_{1} \sigma_{2}) \Lambda (1 - \Lambda)]\}  \label{eq20}  
\end{eqnarray}
with

\begin{equation}
f_{i \alpha \beta} = k_{i \alpha} \epsilon_{i \beta} - \epsilon_{i \alpha} k_{i \beta}   \label{eq20a}
\end{equation}
Equation (\ref{eq20}) can be shown to be equivalent to the result obtained in [22], up to an unphysical overall minus sign. To see the equivalence of the two formulas one would have to transform the variable $s$ according to

\begin{eqnarray}
s & \rightarrow & s \{ e^{2} (q \cdot \bar{F} \cdot \bar{F} \cdot q) \, [ 2 \sigma_{1} \sigma_{2} (1 - 3 \sigma_{1} - 3 \sigma_{2} + 2 \sigma_{1}^{2} + 2 \sigma_{2}^{2} + 3 \sigma_{1} \sigma_{2}) \Lambda (1 - \Lambda)  \nonumber \\
& & \hspace{.3cm} \mbox{} - (\sigma_{1} - \sigma_{1}^{2})^{2} \Lambda^{2} - (\sigma_{2} - \sigma_{2}^{2})^{2} (1 - \Lambda)^{2} ] \}^{1/3}
\end{eqnarray}
perform a Wick-rotation in $s$ and change from Euclidian to Minkowski space.

\section{Remarks about the case $\{F_{\mu \nu} \neq 0, \hspace{.3cm} T, \mu \neq 0\}$}
The finite temperature [25,26,33] and density case even without background field raises some new questions and we first treat this case. Finite temperature is usually taken into account on the world-line in the context of the imaginary time formalism. Thus the time component becomes a circle with circumference $\beta=\frac{1}{T}$. Under this topological constraint the pathintegral in (\ref{eq4}), which is over closed paths, can be written as a sum over pathintegrals without the topological constraint but each one over paths where the endpoint differs from the starting point by a multiple of $\beta$ in the time component [23,24]. This takes into account the different windings of the closed paths on the time-circle. In order to get the statistics right the sum has to be alternating\footnote{Note that this is not in contradiction to the periodicity of $\psi(\tau)$ on the world-line since $\psi(\tau)$ is a world-line and not a space-time fermion.}. That means, if $P^{\beta}_{xy}$ is the pathintegral over paths from $y$ to $x$ under the topological constraint $x_{4} = x_{4} + n \beta$ and $P^{\infty}_{xy}$ is the one without the topological constraint, we have:

\begin{equation}
P^{\beta}_{xy} = \sum_{n=-\infty}^{\infty} (-1)^{n} P^{\infty}_{(\vec{x}, x_{4} + n \beta)y}   \label{eq21}
\end{equation}
Thus we are led to calculate the following expression:

\begin{eqnarray}
{\cal M} & = & - 4 m \lambda' e^{2} \epsilon_{\mu \nu \alpha \gamma} \epsilon_{1 \mu} \epsilon_{2 \nu} k_{1 \alpha} k_{2 \gamma} \int_{0}^{\infty} \frac{ds}{s} \sum_{n=-\infty}^{\infty} \int_{x \rightarrow (\vec{x}, x_{4} + n \beta)} {\cal  D}x \int_{P} {\cal D} \tilde{\psi}   \label{eq22} \\
 & & \int_{0}^{s} d\tau_{1} d\tau_{2} d\tau_{3} e^{i k_{1} x(\tau_{1}) + i k_{2} x(\tau_{2}) + i q x(\tau_{3})} e^{-m^{2} s} \exp\{ -\int_{0}^{s} d\tau \left( \frac{\dot{x}^{2}}{4} + \frac{1}{2} \tilde{\psi} \cdot \dot{\tilde{\psi}} \right) \}  \nonumber
\end{eqnarray}
where we have already performed the integration over the fermionic zero modes. 
We now have to split the path into a linear and a periodic part and separate  the center of mass 

\begin{equation}
x(\tau) \rightarrow x^{(0)} + n \beta \frac{\tau}{s} e_{4} + y(\tau)  \label{eq22a}
\end{equation}
where $e_{4}$ is a unit vector in the direction of the fourth (i.e. time) component. The remaining pathintegrals run over periodic paths with center of mass zero so that the usual Green functions can be used for the Wick contractions (this is true also for the fermionic pathintegrals). The result we get is:

\begin{eqnarray}
{\cal M} & = & - 4 m \lambda' e^{2} \epsilon_{\mu \nu \alpha \gamma} \epsilon_{1 \mu} \epsilon_{2 \nu} k_{1 \alpha} k_{2 \gamma} \, (2 \pi)^{4} \delta(\vec{k}_{1} + \vec{k}_{2} + \vec{q}) \, \delta_{\omega_{k_{1}} + \omega_{k_{2}} + \omega_{q}}  \nonumber \\
& & \int_{0}^{\infty} ds \, s^{2} \int_{0}^{1} d\sigma_{1} d\sigma_{2} \sum_{n=-\infty}^{\infty} (-1)^{n} e^{i (\omega_{k_{1}} \sigma_{1} + \omega_{k_{2}} \sigma_{2}) n \beta} e^{-\frac{n^{2} \beta^{2}}{4 s}} e^{-m^{2} s} (4 \pi s)^{-2}  \nonumber \\
& & \exp\{s (k_{1} \cdot k_{2} \tilde{G}_{B12} + k_{1} \cdot q \tilde{G}_{B1} + k_{2} \cdot q \tilde{G}_{B2})\}   \label{eq23}
\end{eqnarray}
where $\tau_{3}$ has been set to zero again, 
 \begin{equation}
\tilde{G}_{Bij} = |\sigma_{i} - \sigma_{j}| - (\sigma_{i} - \sigma_{j})^{2} \hspace{.2cm} , \hspace{.2cm} \tilde{G}_{Bi} = \sigma_{i} - \sigma_{i}^{2}  \label{eq23a}
\end{equation}
and $\omega_{k}$ are the external bosonic Matsubara frequencies. We now perform a Jakobi-transformation 

\begin{eqnarray}
\lefteqn{\sum_{n=-\infty}^{\infty} (-1)^{n} \exp\{i \pi [\tau n^{2} + 2 z n ]\} = }  \label{eq24} \\
 & = &  (-i \tau)^{-1/2} e^{-\frac{i \pi z^{2}}{\tau}} \sum_{n=-\infty}^{\infty} \exp\left\{i \pi \left[ -\frac{1}{\tau} \left( n + \frac{1}{2} \right)^{2} + 2 \frac{z}{\tau} \left( n + \frac{1}{2} \right) \right] \right\}   \nonumber
\end{eqnarray}
and get\footnote{Because of the analytical properties of the integrand it is possible to change the order of the s-integration and the summation. This is also confirmed by formula (\ref{eq25c}) derived with the Matsubara formalism.}:

\begin{eqnarray}
{\cal M} & = & - \frac{1}{2 \pi^{3/2}} m \lambda' e^{2} \epsilon_{\mu \nu \alpha \gamma} \epsilon_{1 \mu} \epsilon_{2 \nu} k_{1 \alpha} k_{2 \gamma} (2 \pi)^{4} \, \delta(\vec{k}_{1} + \vec{k}_{2} + \vec{q}) \, \delta_{\omega_{k_{1}} + \omega_{k_{2}} + \omega_{q}}  \nonumber \\
& & \int_{0}^{1} d\sigma_{1} d\sigma_{2} \, \beta^{-1} \sum_{n = - \infty}^{\infty} \int_{0}^{\infty} ds \, s^{1/2}  \exp\{-s [m^2 - k_{1} \cdot k_{2} \tilde{G}_{B12} - k_{1} \cdot q \tilde{G}_{B1} \nonumber \\
& &  \mbox{} - k_{2} \cdot q \tilde{G}_{B2} + (\omega_{n}^{-} - \omega_{k_{1}} \sigma_{1} - \omega_{k_{2}} \sigma_{2})^{2}]\}  \label{eq25}  \\
& = & - \frac{1}{2 \pi^{3/2}} m \lambda' e^{2} \epsilon_{\mu \nu \alpha \gamma} \epsilon_{1 \mu} \epsilon_{2 \nu} k_{1 \alpha} k_{2 \gamma} (2 \pi )^{4} \, \delta(\vec{k}_{1} + \vec{k}_{2} + \vec{q}) \, \delta_{\omega_{k_{1}} + \omega_{k_{2}} + \omega_{q}}  \nonumber \\
& & \beta^{-1} \sum_{n = - \infty}^{\infty} \int_{0}^{\infty} ds \, s^{1/2} \left[ \int_{0}^{1} d\sigma_{2} \int_{0}^{1-\sigma_{2}} d\sigma_{1} \exp\{ -s [m^{2} + k_{1}^{2} \sigma_{1}(1-\sigma_{1}) \right. \nonumber \\
& & \hspace{1cm} \mbox{} + k_{2}^{2} \sigma_{2}(1-\sigma_{2}) + 2 k_{1} k_{2} \sigma_{1} \sigma_{2} + (\omega_{n}^{-} - \omega_{k_{1}} (1 - \sigma_{1}) - \omega_{k_{2}} \sigma_{2})^{2} ] \}  \nonumber \\
& & \hspace{1cm} \mbox{} + \int_{0}^{1} d\sigma_{1} \int_{0}^{1-\sigma_{1}} d\sigma_{2} \exp\{ -s [m^{2} + k_{1}^{2} \sigma_{1}(1-\sigma_{1}) + k_{2}^{2} \sigma_{2}(1-\sigma_{2})  \nonumber \\
& &  \hspace{1cm} \left. \mbox{} + 2 k_{1} k_{2} \sigma_{1} \sigma_{2} + (\omega_{n}^{-} - \omega_{k_{1}} \sigma_{1} - \omega_{k_{2}} (1 - \sigma_{2}))^{2} ] \} \right]  \label{eq25b}
\end{eqnarray}
where the ''$-$'' in $\omega_{n}^{-}$ characterizes the Matsubara frequencies as fermionic ones. We can now introduce the chemical potential via $\omega_{n}^{-} \rightarrow \omega_{n}^{-} - i \mu$.  We convert the sum in (\ref{eq25}) into contour integrals via:

\begin{eqnarray}
\lefteqn{\beta^{-1} \sum_{n=-\infty}^{\infty} f(i\omega_{n}^{-} + \mu) = } \nonumber \\
& = & \frac{1}{2 \pi i} \left[ \int_{-i \infty + \mu - \epsilon}^{i \infty + \mu - \epsilon} d\omega \, \frac{f(\omega)}{e^{\beta(\omega - \mu)} + 1} - \int_{-i \infty + \mu + \epsilon}^{i \infty + \mu + \epsilon} d\omega \, \frac{f(\omega)}{e^{\beta(\omega - \mu)} + 1} \right] \nonumber \\
& = & -\frac{1}{2 \pi i} \left[ \int_{-i \infty + \mu + \epsilon}^{i \infty + \mu + \epsilon} d\omega \, \frac{f(\omega)}{e^{\beta(\omega - \mu)} + 1} + \int_{-i \infty + \mu - \epsilon}^{i \infty + \mu - \epsilon} d\omega \, \frac{f(\omega)}{e^{\beta(\mu - \omega)} + 1}  \right. \nonumber \\
& & \mbox{} \left. - \int_{-i \infty}^{i \infty} d\omega \, f(\omega) - \oint_{C^{'}} d\omega \, f(\omega) \right]  \label{eq25e}
\end{eqnarray}

\begin{figure}[h]
\begin{center}
\begin{picture}(300,140)(0,0)
  \LongArrow(100,70)(300,70)
  \LongArrow(100,70)(0,70)
  \ArrowLine(100,140)(100,70)
  \ArrowLine(100,70)(100,0)
  \put(80,139){$i \infty$}
  \put(76,0){$-i \infty$}
  \put(90,75){$0$}
  \put(205,75){$\mu$}
  \ArrowLine(200,0)(200,70)
  \ArrowLine(200,70)(200,140)
  \ArrowLine(200,140)(100,140)
  \ArrowLine(100,0)(200,0)
  \put(290,130){$\omega$}
  \EBox(285,125)(300,140)
\end{picture}
\caption{Contour $C^{'}$}   \label{fig2}
\end{center}
\end{figure}
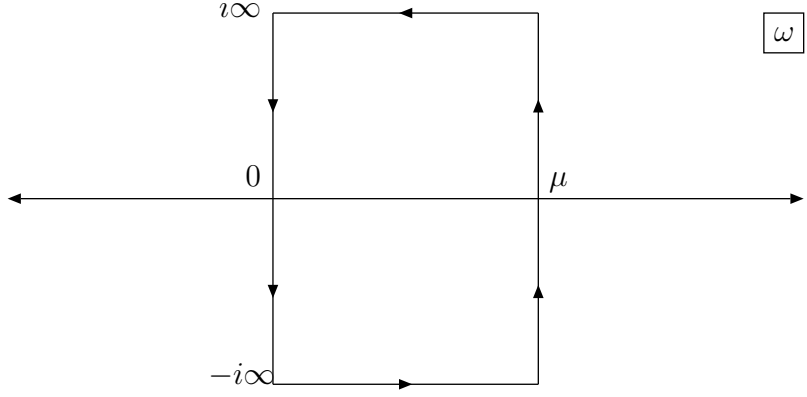

For the case $\mu > 0$ the contour $C^{'}$ is depicted in (fig. \ref{fig2}) and we have assumed that $f(\omega)$ is analytic in a neighbourhood of the line $\Re(\omega)=\mu$ and $\lim_{R \rightarrow \pm  \infty} f(a + i R) = 0$ for $0 < a < \mu$ (which is true for the case at hand). 

In the following we consider only the case $|\mu|<m$. Then our $f(\omega)$ is analytic inside the contour $C^{'}$ and therefore the last term of (\ref{eq25e}) does not contribute. 
  
We are thus led to calculate the following expression for the temperature dependent part:

\begin{eqnarray}
{\cal{M}}^{\beta} & = & \frac{1}{4 \pi} m \lambda^{'} e^{2} \epsilon_{\mu \nu \alpha \gamma} \epsilon_{1 \mu} \epsilon_{2 \nu} k_{1 \alpha} k_{2 \gamma} (2 \pi)^{4} \, \delta(\vec{k}_{1} + \vec{k}_{2} + \vec{q}) \, \delta_{\omega_{k_{1}} + \omega_{k_{2}} + \omega_{q}}  \nonumber \\
& & \int_{0}^{1} d\sigma_{1} d\sigma_{2} \left[ \frac{1}{2 \pi i}\int_{-i \infty + \mu + \epsilon}^{i \infty + \mu + \epsilon} d\omega \frac{(A - (\omega - i B)^{2})^{-3/2}}{e^{\beta (\omega - \mu)}+1} \right.  \nonumber \\
& & \mbox{} \left. + \frac{1}{2 \pi i} \int_{-i \infty + \mu - \epsilon}^{i \infty +\mu - \epsilon} d\omega \frac{(A - (\omega - i B)^{2})^{-3/2}}{e^{- \beta (\omega - \mu)}+1} \right]  \label{eq25k}  
\end{eqnarray}
with

\begin{eqnarray}
A & = & m^{2} - k_{1} \cdot k_{2} \tilde{G}_{B12} - k_{1} \cdot q \tilde{G}_{B1} - k_{2} \cdot q \tilde{G}_{B2} \nonumber \\
B & = & \omega_{k_{1}} \sigma_{1} + \omega_{k_{2}} \sigma_{2} \label{eq25i}
\end{eqnarray}
We now have to deform the $\omega$-contour to the real axis. In doing so we have to be careful because of the squareroot cut in the integrand. The details of the calculation can be found in appendix C and here we just give the result:

\begin{eqnarray}
{\cal M}^{\beta} & = & \frac{1}{4 \pi^{2}} m \lambda' e^{2} \epsilon_{\mu \nu \alpha \gamma} \epsilon_{1 \mu} \epsilon_{2 \nu} k_{1 \alpha} k_{2 \gamma} (2 \pi)^{4} \, \delta(\vec{k}_{1} + \vec{k}_{2} + \vec{q}) \, \delta_{\omega_{k_{1}} + \omega_{k_{2}} + \omega_{q}} \int_{0}^{1} d\sigma_{1} d\sigma_{2}  \nonumber \\
& & \int_{0}^{\infty} dp \, \left[ \frac{A^{-1} (p^{2}+1)^{-3/2}}{e^{\beta (\sqrt{A} \sqrt{p^{2}+1} - \mu +i B)} + 1} + \frac{\beta A^{-1/2} (p^{2}+1)^{-1} e^{\beta (\sqrt{A} \sqrt{p^{2}+1} -\mu +i B)}}{(e^{\beta (\sqrt{A} \sqrt{p^{2}+1} - \mu + i B)} + 1)^{2}} \right.  \nonumber \\
& & \hspace{-.2cm} \left. \mbox{} + \frac{A^{-1} (p^{2}+1)^{-3/2}}{e^{-\beta (-\sqrt{A} \sqrt{p^{2}+1} - \mu +i B)} + 1} + \frac{\beta A^{-1/2} (p^{2}+1)^{-1} e^{-\beta (-\sqrt{A} \sqrt{p^{2}+1} -\mu +i B)}}{(e^{-\beta (-\sqrt{A} \sqrt{p^{2}+1} - \mu + i B)} + 1)^{2}}  \right]  \nonumber \\
& &  \label{eq25h}
\end{eqnarray} 
In appendix C we derive this result also with a different method. 

Decomposing $(p^{2}+1)^{-1} = -p^{2}/(p^{2}+1)^{-1} + 1$ in the second and fourth term and performing a partial integration in the $-p^{2}/(p^{2}+1)^{-1}$ part one arrives at a simpler expression\footnote{This was suggested by H. Sato.}:

\begin{eqnarray}
{\cal M}^{\beta} & = & \frac{1}{4 \pi^{2}} m \lambda' e^{2} \epsilon_{\mu \nu \alpha \gamma} \epsilon_{1 \mu} \epsilon_{2 \nu} k_{1 \alpha} k_{2 \gamma} (2 \pi)^{4} \, \delta(\vec{k}_{1} + \vec{k}_{2} + \vec{q}) \, \delta_{\omega_{k_{1}} + \omega_{k_{2}} + \omega_{q}} \int_{0}^{1} d\sigma_{1} d\sigma_{2} \nonumber \\
& & \hspace{-.6cm} \int_{0}^{\infty} dp \, \beta A^{-1/2} \left[ \frac{e^{\beta (\sqrt{A} \sqrt{p^{2}+1} -\mu +i B)}}{(e^{\beta (\sqrt{A} \sqrt{p^{2}+1} - \mu + i B)} + 1)^{2}} + \frac{e^{-\beta (-\sqrt{A} \sqrt{p^{2}+1} -\mu +i B)}}{(e^{-\beta (-\sqrt{A} \sqrt{p^{2}+1} - \mu + i B)} + 1)^{2}}  \right] \nonumber \\
& & \label{eq25h2} 
\end{eqnarray}

 The general $\sigma$-integration is not possible here. A similar problem has been encountered during the investigation of a scalar quantum field theory at two loop level [31]. One could try to specialize to the case where the rest frame of the axion is identical with the frame of the heat bath and perform the analytical continuation in the external Matsubara frequencies ($i \omega_{k_{j}} \rightarrow (k_{j})_{0} - i \epsilon_{j} \hspace{.2cm} , \hspace{.2cm} j=1,2$ , see e.g. [27] and for a general treatment of the relationship between imaginary-time and real-time finite temperature field theory also [28-30]). As pointed out in [26] the result should coincide with the corresponding calculation in the real-time formalism (see e.g. [25]), if we assume $M<<m$, where $M$ is the axion mass. The examination of the analytical continuation has not been finished yet. Some details can be found in appendix D.

It is possible to show that our expression agrees with the one one gets with the usual Matsubara-Feynman-rules, if one introduces Schwinger parameters and integrates out the spatial momenta. In order to get (\ref{eq25}) one has to use the relationship between Feynman parameters and world-line times (see (fig. \ref{fig3})):

\begin{equation}
\alpha_{1} = \tau_{1} = s \sigma_{1}  \hspace{.4cm} , \hspace{.4cm} \alpha_{2} = \tau_{2} - \tau_{1}  = s (\sigma_{2} - \sigma_{1}) \hspace{.4cm} , \hspace{.4cm} \alpha_{3} = s - \tau_{2} = s (1 - \sigma_{2})  \label{eq25a}
\end{equation}
In detail we have:

\begin{figure}[t]
\begin{center}
\begin{picture}(350,150)(0,0)
  \DashArrowLine(10,75)(60,75)5
  \put(1,75){$q$}
  \put(34,82){$0 \! = \! \tau_{3}$}
  \ArrowLine(60,75)(110,125)
  \put(62,110){$p, \, \alpha_{1}$}
  \ArrowLine(110,25)(60,75)
  \put(34,46){$p-q, \, \alpha_{3}$}
  \ArrowLine(110,125)(110,25)
  \put(120,75){$p+k_{1}, \, \alpha_{2}$}
  \Photon(110,125)(160,125)35
  \put(105,134){$\tau_{1}$}
  \Photon(110,25)(160,25)35
  \put(105,17){$\tau_{2}$}
  \put(135,135){$k_{1},\epsilon_{\mu}$}
  \put(135,10){$k_{2},\epsilon_{\nu}$}
  \put(210,75){$+$}
  \put(250,75){$(k_{1},\epsilon_{\mu}) \longleftrightarrow (k_{2},\epsilon_{\nu})$}
\end{picture}
\caption{Relationship between Feynman parameters and world-line times}
\label{fig3}
\end{center}
\end{figure}
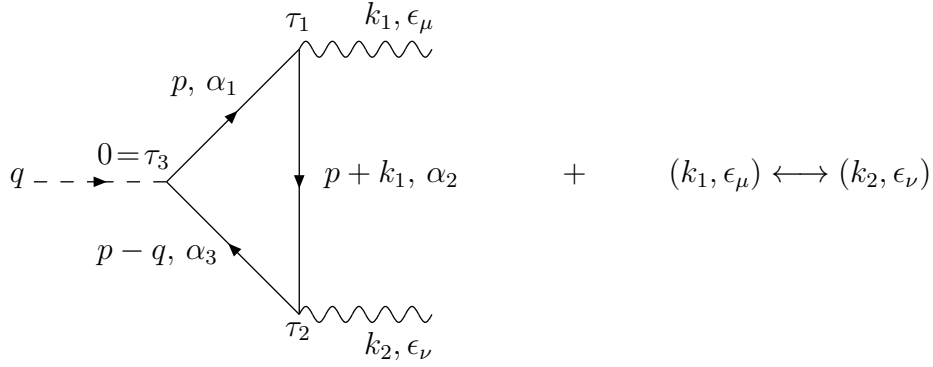

\begin{eqnarray}
{\cal M} & = & -\epsilon_{1 \mu} \epsilon_{2 \nu} \beta^{-1} \sum_{n=-\infty}^{\infty} \int \frac{d^{3}p}{(2 \pi)^{3}} \mbox{tr} \left[ \frac{m - (p \hspace{-.17cm} / \hspace{.1cm} + k_{1} \hspace{-.35cm}/ \hspace{.2cm})}{(\omega_{n}^{-} + \omega_{k_{1}})^{2} + (\vec{p} + \vec{k_{1}})^{2} + m^{2}} \gamma_{\mu} \right. \nonumber \\
& & \hspace{2.5cm} \left. \frac{m - p \hspace{-.17cm}/}{(\omega_{n}^{-})^{2} + \vec{p}^{2} + m^{2}} \gamma_{5} \frac{m - (p \hspace{-.17cm} / \hspace{.1cm} - q \hspace{-.18cm}/ \hspace{.1cm})}{(\omega_{n}^{-} - \omega_{q})^{2} + (\vec{p} - \vec{q})^{2} + m^{2}} \gamma_{\nu} \right]  \nonumber \\
& = &  - 4 m \epsilon_{\mu \nu \alpha \gamma} \epsilon_{1 \mu} \epsilon_{2 \nu} k_{1 \alpha} k_{2 \gamma} \beta^{-1} \sum_{n=-\infty}^{\infty} \int \frac{d^{3}p}{(2 \pi)^{3}} \int_{0}^{\infty} d\alpha_{1} d\alpha_{2} d\alpha_{3} \,  \nonumber  \\
& & \exp\{ -\alpha_{1} ((\omega_{n}^{-})^{2} + \vec{p}^{2} + m^{2}) - \alpha_{2}((\omega_{n}^{-} + \omega_{k_{1}})^{2} + (\vec{p} + \vec{k}_{1})^{2} + m^{2})  \nonumber \\
& &  \hspace{.75cm} \mbox{} - \alpha_{3}((\omega_{n}^{-} - \omega_{q})^{2} + (\vec{p} - \vec{q})^{2} + m^{2})\}  \nonumber \\ 
& = & - 4 m \epsilon_{\mu \nu \alpha \gamma} \epsilon_{1 \mu} \epsilon_{2 \nu} k_{1 \alpha} k_{2 \gamma}  \beta^{-1} \sum_{n=-\infty}^{\infty} \int_{0}^{\infty} d\alpha_{1} d\alpha_{2} d\alpha_{3} \, (4 \pi (\alpha_{1} + \alpha_{2} + \alpha_{3}))^{-3/2}  \nonumber \\
& & \exp\{- m^{2} (\alpha_{1} + \alpha_{2} + \alpha_{3}) - (\omega_{n}^{-})^{2} (\alpha_{1} + \alpha_{2} + \alpha_{3}) - \omega_{k_{1}}^{2} \alpha_{2} - \omega_{q}^{2} \alpha_{3}  \nonumber \\
& & \hspace{.75cm} \mbox{} + \omega_{n}^{-} (2 \omega_{q} \alpha_{3} - 2 \omega_{k_{1}} \alpha_{2}) - \vec{q}^{2} \alpha_{3} - \vec{k}_{1}^{2} \alpha_{2} + \frac{(\vec{q} \alpha_{3} - \vec{k}_{1} \alpha_{2})^{2}}{\alpha_{1} + \alpha_{2} + \alpha_{3}} \}  \nonumber \\ 
& \stackrel{(\ref{eq25a})}{=} & - 4 m \epsilon_{\mu \nu \alpha \gamma} \epsilon_{1 \mu} \epsilon_{2 \nu} k_{1 \alpha} k_{2 \gamma}  \beta^{-1} \sum_{n=-\infty}^{\infty} \int_{0}^{\infty} ds (4 \pi s)^{-3/2} s^{2} \int_{0}^{1} d\sigma_{2} \int_{0}^{\sigma_{2}} d\sigma_{1} \nonumber \\
& & \exp\{ -s [m^{2} + (\omega_{n}^{-} + \omega_{k_{1}} (1 - \sigma_{1}) + \omega_{k_{2}} (1 - \sigma_{2}))^{2} + k_{1}^{2} \sigma_{1} (1-\sigma_{1})  \nonumber \\
& & \mbox{} + k_{2}^{2} \sigma_{2} (1 - \sigma_{2}) + 2 k_{1} k_{2} \sigma_{1} (1 - \sigma_{2}) ] \}  \label{eq25c}
\end{eqnarray}
where we have used $\mbox{tr}(\gamma_{5} \gamma_{\mu} \gamma_{\nu} \gamma_{\alpha} \gamma_{\beta}) = 4 \epsilon_{\mu \nu \alpha \beta}$
with the Euclidian conventions $\epsilon_{1234} = 1$, $\gamma_{5} = - \gamma_{1} \gamma_{2} \gamma_{3} \gamma_{4}$, $\gamma_{4} = i \gamma_{0}$. In order to get the first term in the square brackets of (\ref{eq25b}) one has to transform $\sigma_{2} \rightarrow 1 - \sigma_{2}$ and $n \rightarrow -n$ in (\ref{eq25c}). The second term in the square brackets of (\ref{eq25b}) is derived by interchange of $\tau_{1}$ and $\tau_{2}$.
 
With the same technique it is possible to treat the case with a constant background field. The formula we derive along the lines outlined above is:

\begin{eqnarray}
{\cal M} & = & \frac{\lambda' m}{8 \pi^{3/2}} (2 \pi)^{4} \delta(\vec{k}_{1} + \vec{k}_{2} + \vec{q}) \, \delta_{\omega_{k_{1}} + \omega_{k_{2}} + \omega_{q}} \int_{0}^{\infty} ds \int_{0}^{1} d\sigma_{1} d\sigma_{2} d\sigma_{3} \, e^{- m^{2} s}  \nonumber \\
& & \beta^{-1} \sum_{n=-\infty}^{\infty} \exp\{ k_{1} \cdot {\cal{G}}(\tau_{1},\tau_{2}) \cdot k_{2} + k_{1} \cdot {\cal{G}}(\tau_{1},\tau_{3}) \cdot q + k_{2} \cdot {\cal{G}}(\tau_{2},\tau_{3}) \cdot q \}  \nonumber \\
& & \exp\{ - s (\omega_{n}^{-} - \omega_{k_{1}} \sigma_{1} - \omega_{k_{2}} \sigma_{2} - \omega_{q} \sigma_{3})^{2} \} \left\{\tilde{{\cal M}}_{0} + \tilde{{\cal M}}_{1} + \tilde{{\cal M}}_{2} \right\}   \label{eq26}
\end{eqnarray}
with

\begin{eqnarray}
\tilde{{\cal M}}_{0} & = & 2 e^{2} s^{1/2} (\hat{f}_{1 \mu \nu} f_{2 \mu \nu})   \label{eq27} \\
\tilde{{\cal M}}_{1} & = & 4 i e^{3} s^{3/2} \left[ \frac{1}{2} (f_{2 \mu \nu} \hat{F}_{\mu \nu}) \left((\epsilon_{1} \cdot \dot{{\cal{G}}}(\tau_{1},\tau_{2}) \cdot k_{2}) + (\epsilon_{1} \cdot \dot{{\cal{G}}}(\tau_{1},\tau_{3}) \cdot{q}) \right) \right. \nonumber \\
& & \mbox{} - \frac{1}{2}(k_{1} \cdot \hat{F} \cdot k_{2})(\epsilon_{1} \cdot \dot{{\cal{G}}}(\tau_{1},\tau_{2}) \cdot \epsilon_{2})  \nonumber \\
& & \mbox{} + (k_{1} \cdot \hat{F} \cdot \epsilon_{2})(\epsilon_{1} \cdot \dot{{\cal{G}}}(\tau_{1},\tau_{2}) \cdot k_{2}) \nonumber \\
& & \mbox{} + (\omega_{n}^{-} - \omega_{k_{1}} \sigma_{1} - \omega_{k_{2}} \sigma_{2} - \omega_{q} \sigma_{3}) (\epsilon_{1})_{4} (f_{2 \mu \nu} \hat{F}_{\mu \nu})  \nonumber \\
& & \mbox{} \left. - \frac{1}{2} (\epsilon_{1} \cdot \hat{F} \cdot \epsilon_{2})(k_{1} \cdot \dot{{\cal{G}}}(\tau_{1},\tau_{2}) \cdot k_{2}) \right] \hspace{.2cm} + \hspace{.2cm} (\epsilon_{1}, k_{1}) \leftrightarrow (\epsilon_{2}, k_{2})   \label{eq28} \\
\tilde{{\cal M}}_{2} & = & - e^{4} s^{5/2} (\hat{F}_{\mu \nu} F_{\mu \nu}) \left[ \frac{1}{2} (\epsilon_{1} \cdot \dot{{\cal{G}}}(\tau_{1},\tau_{2}) \cdot \epsilon_{2})(k_{1} \cdot \dot{{\cal{G}}}(\tau_{1},\tau_{3}) \cdot q) \right. \nonumber \\
& & \mbox{} + (\epsilon_{1} \cdot \dot{{\cal{G}}}(\tau_{1},\tau_{3}) \cdot q)(\epsilon_{2} \cdot \dot{{\cal{G}}}(\tau_{2},\tau_{1}) \cdot k_{1})  \nonumber \\
& & \mbox{} + \left\{ -\frac{1}{s} + 2 (\omega_{n}^{-} - \omega_{k_{1}} \sigma_{1} - \omega_{k_{2}} \sigma_{2} - \omega_{q} \sigma_{3})^{2} \right\} (\epsilon_{1})_{4} (\epsilon_{2})_{4}  \nonumber \\
& & \mbox{} - 2 (\omega_{n}^{-} - \omega_{k_{1}} \sigma_{1} - \omega_{k_{2}} \sigma_{2} - \omega_{q} \sigma_{3}) \left\{\frac{1}{2}(k_{1})_{4}(\epsilon_{1} \cdot \dot{{\cal{G}}}(\tau_{1},\tau_{2}) \cdot \epsilon_{2}) \right.  \nonumber \\
& & \left. \mbox{} - (\epsilon_{1})_{4} \left((\epsilon_{2} \cdot \dot{{\cal{G}}}(\tau_{2},\tau_{1}) \cdot k_{1}) + (\epsilon_{2} \cdot \dot{{\cal{G}}}(\tau_{2},\tau_{3}) \cdot q) \right) \right\} \nonumber \\
& & \left. \mbox{} + \frac{1}{2} (\epsilon_{1} \cdot \dot{{\cal{G}}}(\tau_{1},\tau_{3}) \cdot q)(\epsilon_{2} \cdot \dot{{\cal{G}}}(\tau_{2},\tau_{3}) \cdot q) \right]  \nonumber \\
& & \mbox{} + (\epsilon_{1}, k_{1}) \leftrightarrow (\epsilon_{2}, k_{2})   \label{eq29}
\end{eqnarray}
where we have used (\ref{eq20a}) and $\hat{f}_{\mu \nu} = \frac{1}{2} \epsilon_{\mu \nu \alpha \gamma} f_{\alpha \gamma}$. $\dot{{\cal{G}}}(\tau_{i},\tau_{j})$ is of course understood as $\dot{{\cal{G}}}(s \sigma_{i},s \sigma_{j})$. It is again possible to fix $\sigma_{3} = 0$. $(\epsilon_{i})_{4}$ and $(k_{i})_{4}$ mean the fourth (i.e. timelike) component of the corresponding vector in brackets.  

We have seen that the world-line formalism provides a powerful tool in treating problems with a constant electromagnetic background field. The examination of temperature effects still suffers from the problem, that the occuring integrals are not solvable in general and that the result has to be continued analytically. These problems have also been discussed in [31], where it was shown however, that the world-line method offers an illustrative way of finding the divergent parts of the $\phi^{3}_{6}$-two-loop-amplitude at finite temperature. It is too early to draw a final conclusion about the usefulness of the world-line method in finite temperature field theory. It might also be possible to combine the real time formalism with the world-line method, so that no problems of analytical continuation occur. Here investigations are in progress.

\section*{Acknowledgements}
We would like to thank H.~Nachbagauer and H.~Sato for very useful discussions. This work was supported in part by the TMR network {\em Finite Temperature Phase Transitions in Particle Physics}, EU contract no. ERBFMRXCT97-0122.

\appendix
\section{Green functions and determinants in a constant background field}

\setcounter{equation}{0}
\renewcommand{\theequation}{\Alph{section}.\arabic{equation}}

In a constant background the defining equation for the bosonic Green function becomes

\begin{equation}
\frac{1}{2} \left( \delta_{\mu \lambda} \frac{\partial^{2}}{\partial \tau^{2}} - 2 i e \bar{F}_{\mu \lambda} \frac{\partial}{\partial \tau} \right) {\cal{G}}_{B}^{\lambda \nu}(\tau_{1},\tau_{2}) = \delta_{\mu \nu} \left(\delta(\tau_{1} - \tau_{2}) - \frac{1}{s} \right)   \label{eqa1}
\end{equation}
In the case at hand the fermions obey periodic boundary conditions. Thus we have [16]:

\begin{eqnarray}
{\cal{G}}_{B}(\tau_{1}, \tau_{2}) & = & \frac{1}{2(e\bar{F})^{2}} \left( \frac{e\bar{F}}{\sin(e\bar{F}s)} e^{-i e \bar{F} s \dot{G}_{B12}} + i e \bar{F} \dot{G}_{B12} - \frac{1}{s} \right)    \label{eqa2} \\
{\cal{G}}_{F}(\tau_{1}, \tau_{2}) & = & \dot{{\cal{G}}}_{B}(\tau_{1}, \tau_{2}) \nonumber \\
 & = & \frac{i}{e\bar{F}} \left( \frac{e\bar{F}}{\sin(e \bar{F} s)} e^{-ie\bar{F}s\dot{G}_{B12}} - \frac{1}{s} \right)    \label{eqa3}
\end{eqnarray}
where 

\begin{equation}
G_{B12} = |\tau_{1} - \tau_{2}| - \frac{(\tau_{1} - \tau_{2})^{2}}{s}  \label{eqa3a}
\end{equation}
is the well known bosonic Green function on the circle without background field.

The first few terms of the Taylor expansion in F for the Green functions are:

\begin{eqnarray}
{\cal{G}}_{B12} & = & G_{B12} - \frac{i}{3} \dot{G}_{B12} G_{B12} s e \bar{F} + \frac{s}{3} G^{2}_{B12} (e\bar{F})^{2} + {\cal{O}}(\bar{F}^{3})   \label{eqa4}  \\
{\cal{G}}_{F12} & = & \dot{{\cal{G}}}_{B12} = \dot{G}_{B12} + 2 i (G_{B12} - \frac{s}{6}) e \bar{F} + \frac{2}{3} \dot{G}_{B12} G_{B12} s (e \bar{F})^{2} + {\cal{O}}(\bar{F}^{3})  \nonumber \\
& &   \label{eqa5}
\end{eqnarray}
where in the expression for ${\cal{G}}_{B12}$ the constant coincidence limit has already been subtracted (which is possible because of momentum conservation). 

The free pathintegrals in a constant background and with the fermions obeying periodic boundary conditions are:

\begin{eqnarray}
\lefteqn{\int_{P} {\cal{D}}x \, \exp \left\{- \frac{1}{4} \int_{0}^{s} d\tau \, (\dot{x}^{2} + 2 i e x_{\mu} \bar{F}_{\mu \nu} \dot{x}_{\nu}) \right\} = } \nonumber \\ 
& = & \underbrace{\det \, \! ^{'-\frac{1}{2}}_{P}[-\partial^{2}_{\tau}]}_{\mbox{\large{$(4 \pi s)^{-2}$}}} \det \, \! ^{'-\frac{1}{2}}_{P}[{\bf I} - 2 i e \bar{F} \partial^{-1}_{\tau}]   \label{eqa6} \\    
\lefteqn{\int_{P} {\cal{D}} \tilde{\psi} \, \exp\left\{ -\frac{1}{2} \int_{0}^{s} d\tau \, \left( \tilde{\psi} \dot{\tilde{\psi}} - 2 i e \tilde{\psi}_{\mu} F_{\mu \nu} \tilde{\psi}_{\nu} \right) \right\} = }  \nonumber  \\
&  = & \underbrace{\det \, \! ^{' \frac{1}{2}}_{P} [\partial]}_{\mbox{\large{$1$}}} \det \, \! ^{' \frac{1}{2}}_{P} [{\bf I} - 2 i e \bar{F} \partial^{-1}_{\tau}]  \label{eqa7}   
\end{eqnarray}
where the prime at the determinant reminds at the fact, that the zeromodes are excluded.

\pagebreak

\section{Derivation of the formulas (\ref{eq19}) and (\ref{eq20})}

\setcounter{equation}{0}

In the case $k_{1} \parallel k_{2} \parallel q$ we have, because of the antisymmetry of $F$ and $\dot{{\cal{G}}}$, that the terms proportional to $k_{1} \cdot \hat{F} \cdot k_{2}$ and $k_{1} \cdot \dot{{\cal{G}}}(\tau_{1}, \tau_{2}) \cdot k_{2}$ do  not contribute (realize, that the order of $\tau_{1}$ and $\tau_{2}$ in the argument of $\dot{{\cal{G}}}$ can be changed because we are integrating both over the whole interval [0,s]; thus (\ref{eq13}) can be applied to see, that the term does not contribute). In order to derive (\ref{eq19}) we take the first term in the square brackets of (\ref{eq17}) as it stands and the last one with $(\epsilon_{1}, k_{1}) \leftrightarrow (\epsilon_{2}, k_{2})$. Using the fact that $k_{1} = - \Lambda q$ and $k_{2} = - (1 - \Lambda) q$ we immediately get:

\begin{eqnarray}
[...] & = & 2 (k_{1} \cdot \hat{F} \cdot \epsilon_{1}) \left\{ (\epsilon_{2} \cdot \dot{{\cal{G}}} (\tau_{2}, \tau_{1}) \cdot k_{1}) + (\epsilon_{2} \cdot \dot{{\cal{G}}} (\tau_{2}, \tau_{3}) \cdot q) \right\}  \nonumber \\
& & \mbox{} + 2 (k_{2} \cdot \hat{F} \cdot \epsilon_{1}) (\epsilon_{2} \cdot \dot{{\cal{G}}} (\tau_{1}, \tau_{2}) \cdot k_{1})  \nonumber \\
& = & 2 \Lambda (1 - \Lambda) (q \cdot \hat{F} \cdot \epsilon_{1}) \left[\epsilon_{2} \cdot \dot{{\cal{G}}}(\tau_{1}, \tau_{2}) \cdot q - \epsilon_{2} \cdot \dot{{\cal{G}}}(\tau_{2}, \tau_{1}) \cdot q \right]  \nonumber \\
& & \mbox{} + \frac{2}{1 - \Lambda} (k_{1} \cdot \hat{F} \cdot \epsilon_{1}) \left[\epsilon_{2} \cdot \left(\dot{{\cal{G}}}(\tau_{2}, \tau_{1}) - \dot{{\cal{G}}}(\tau_{2}, \tau_{3} \right) \cdot k_{2} \right]  \label{eqd1}
\end{eqnarray}
the first summand of (\ref{eqd1}) vanishes upon integration over the $\tau_{i}$ as explained above.

To get from (\ref{eq19}) to (\ref{eq20}) we rescale the $\tau_{i}=s \sigma_{i}$ and set $\sigma_{3} = 0$. Furthermore we realize that in the expansion of $\epsilon_{2} \cdot \left(\dot{{\cal{G}}}(s \sigma_{2},s \sigma_{1}) - \dot{{\cal{G}}}(s \sigma_{2}, 0) \right) \cdot k_{2}$ only the second term of (\ref{eqa5}) contributes. The first one is proportional to the unit matrix and we have $\epsilon_{i} \cdot k_{i} = 0$ because of the Lorentz gauge choice. The third one is odd  under interchange of world-line times. Thus integration of $\sigma_{1}$ and $\sigma_{2}$ over the whole interval $[0,1]$ yields zero. As already stated in the text it is easy to verify that $\bar{F}_{\mu \alpha} \bar{F}_{\alpha \beta} \bar{F}_{\beta \nu} = 0$. 

Moreover in the exponent of $\exp\{\frac{1}{2}\sum_{i \neq j} k_{i} \cdot {\cal{G}} (\tau_{i}, \tau_{j}) \cdot k_{j} \}$ only the third term of (\ref{eqa4}) contributes in the limit of vanishing axion mass. In this case $q^{2} = 0$, so that the first term yields zero and the second term vanishes because of the antisymmetry of $\bar{F}$.

If we now split the range of integration into $(\sigma_{1} \in [0,1], \sigma_{2} \in [0,\sigma_{1}])$ and $(\sigma_{1} \in [0,1], \sigma_{2} \in [\sigma_{1}, 1])$ respectively and perform a change of variables ($\sigma_{1} \rightarrow 1 - \sigma_{1}$ in the first case and $\sigma_{2} \rightarrow 1 - \sigma_{2}$ in the second case) we see that both ranges contribute equally and we end up with formula (\ref{eq20}).

\pagebreak

\section{Derivation of formula (\ref{eq25h})}

\setcounter{equation}{0}

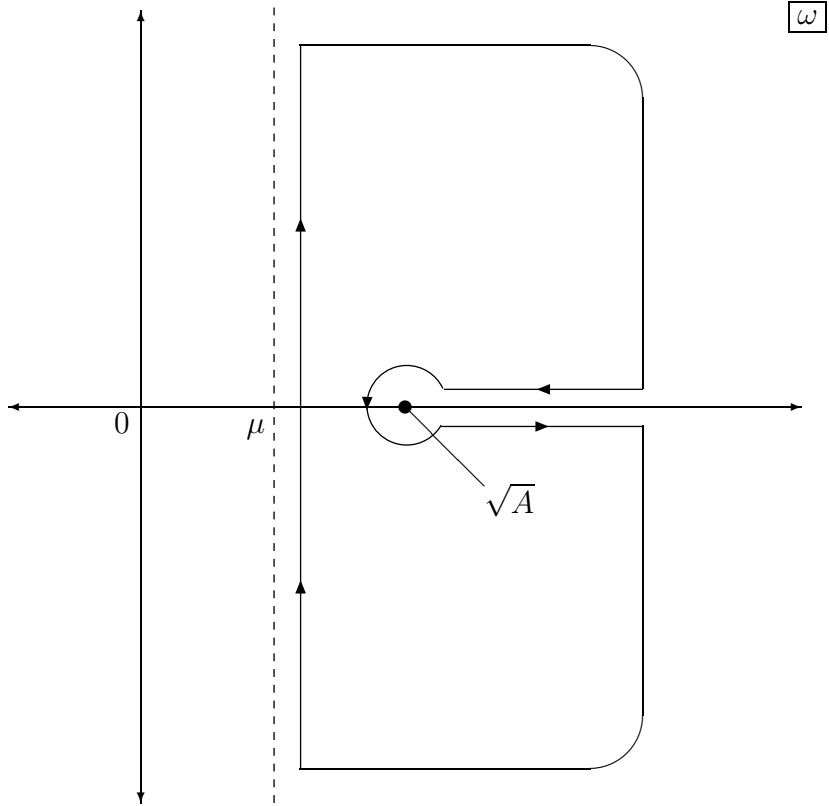
\begin{figure}[h]
\begin{center}
\begin{picture}(350,325)(0,0)
  \put(125,175){\vector(-1,0){100}}
  \put(125,175){\vector(1,0){200}}
  \put(75,175){\vector(0,1){150}}
  \put(75,175){\vector(0,-1){150}}
  \put(65,165){$0$}
  \DashLine(125,25)(125,325)3
  \put(115,165){$\mu$}
  \put(320,320){\fbox{$\omega$}}
  \put(175,175){\circle*{5}}
  \ArrowLine(135,175)(135,311)
  \ArrowLine(135,38)(135,175)
  \put(135,182){\oval(260,260)[tr]}
  \put(135,168){\oval(260,260)[br]}
  \ArrowLine(264,181)(189,181)
  \ArrowLine(188,167)(264,167)
  \ArrowArc(175,175)(15,25,330)
  \thinlines
  \put(175,175){\line(1,-1){30}}
  \put(205,135){$\sqrt{A}$}
\end{picture}
\caption{Deformation of the $\omega$-contour}
\label{fig4}
\end{center}
\end{figure}

Our starting point is equation (\ref{eq25k}). We concentrate on the first of the two integrals, the second one is treated in an analogous way. The aim is to deform the $\omega$-integration-contour to the real axis as depicted in (fig. \ref{fig4}).

First we make a shift in $\omega$ by $i B$ and rescale by $\sqrt{A}^{-1}$. This is possible because $\sqrt{A} \ge m > 0$ for all values of $\sigma_{1/2}$.\footnote{Explicitly we have:
$A = m^{2} + 2 \vec{k}_{1} \cdot \vec{k}_{2} \sigma_{2} (1 - \sigma_{1}) 
+ \vec{k}^{2}_{1} (\sigma_{1} - \sigma_{1}^{2}) + \vec{k}^{2}_{2} (\sigma_{2} - \sigma_{2}^{2}) + 2 \omega_{1} \omega_{2} \sigma_{2} (1 - \sigma_{1}) + \omega^{2}_{1} (\sigma_{1} - \sigma_{1}^{2}) + \omega_{2}^{2} (\sigma_{2} - \sigma_{2}^{2})$ 
for $\sigma_{1} \ge \sigma_{2}$ and a similar expression for $\sigma_{2} \ge \sigma_{1}$ ($\sigma_{1} \leftrightarrow \sigma_{2}, k_{1} \leftrightarrow k_{2}$).}
Thus we get:

\begin{eqnarray}
\lefteqn{\int_{-i \infty + \epsilon +\mu}^{i \infty + \epsilon + \mu} d\omega \, \frac{\left[A - (\omega - i B)^{2} \right]^{-\frac{3}{2}}}{e^{\beta(\omega - \mu)} + 1}}  \nonumber \\
& = & A^{-1} \int_{1 + \delta_{1} + i \delta_{2}}^{\infty + i \delta_{2}} d\omega \frac{\left[ 1 - \omega^{2} \right]^{- \frac{3}{2}}}{e^{\beta (\omega \sqrt{A} - \mu + i B)} + 1} - A^{-1} \int_{1 + \delta_{1} - i \delta_{2}}^{\infty - i \delta_{2}} d\omega \frac{\left[ 1 - \omega^{2} \right]^{- \frac{3}{2}}}{e^{\beta (\omega \sqrt{A} - \mu + i B)} + 1}  \nonumber \\
& &  \mbox{} + A^{-1} i \delta_{1}^{-1/2} \int_{-\pi}^{\pi} d\varphi \, \frac{e^{-\frac{i}{2} \varphi} (2 - \delta_{1} e^{i \varphi})^{-\frac{3}{2}}}{e^{\beta ((1 - \delta_{1} \exp\{i \varphi\}) \sqrt{A} - \mu + i B)} + 1}  \label{eqb1}
\end{eqnarray}
In the last integral we have introduced the variable $z = 1 - \omega = \delta_{1} e^{i \varphi}$. We have also taken the limit $\delta_{2} \rightarrow 0$ here. The range of integration comes from the fact, that we have taken the squareroot cut along the negative real axis. As we are interested in the limit $\delta_{1} \rightarrow 0$ the only contribution from the last integral is 

\begin{equation}
\frac{A^{-1} 2^{-1} i (2 \delta_{1})^{-1/2}}{e^{\beta (\sqrt{A} - \mu + i B)} + 1} \int_{-\pi}^{\pi} d\varphi \, e^{-\frac{i}{2} \varphi} = \frac{A^{-1} 2 i (2 \delta_{1})^{-1/2}}{e^{\beta (\sqrt{A} - \mu + i B)} + 1}  \label{eqb2}
\end{equation}
The first two integrals in (\ref{eqb1}) give the same contribution. If we have a positive imaginary part for $\omega$ the imaginary part of $1-\omega^{2}$ is negative. Therefore we have in the first integral $(1 - \omega^{2})^{-3/2} = \left(e^{-i \pi} (\omega^{2} - 1) \right)^{-3/2} = - i (\omega^{2} - 1)^{-3/2}$ and a relative minus sign in the second one. If we introduce the variable $p^{2} = \omega^{2} - 1$ we get 

\begin{eqnarray}
\lefteqn{-2 i A^{-1} \int_{1 + \delta_{1}}^{\infty} d\omega \frac{\left[\omega^{2} - 1 \right]^{- \frac{3}{2}}}{e^{\beta (\omega \sqrt{A} - \mu + i B)} + 1}} \nonumber \\
& = & -2 i A^{-1} \int_{\sqrt{2 \delta_{1}}}^{\infty} dp \, \frac{p^{-2} (p^{2} + 1)^{-1/2}}{e^{\beta(\sqrt{p^{2} + 1} \sqrt{A} - \mu + i B)} + 1}  \nonumber \\ 
& \stackrel{p.I.}{=} & -2 i A^{-1} \left[ \frac{(2 \delta_{1})^{-1/2}}{e^{\beta(\sqrt{A} - \mu + i B)} + 1} \right.  \label{eqb3}  \\
& & \hspace{-1cm} \mbox{} \left. - \int_{0}^{\infty} dp \, \left( \frac{(p^{2}+1)^{-3/2}}{e^{\beta (\sqrt{A} \sqrt{p^{2}+1} - \mu +i B)} + 1} + \frac{\beta A^{-1/2} (p^{2}+1)^{-1} e^{\beta (\sqrt{A} \sqrt{p^{2}+1} -\mu +i B)}}{(e^{\beta (\sqrt{A} \sqrt{p^{2}+1} - \mu + i B)} + 1)^{2}} \right) \right]  \nonumber
\end{eqnarray}  
where we have omitted terms which vanish in the limit $\delta_{1} \rightarrow 0$. We see that the singular part cancels the one from (\ref{eqb2}) and we are left with the first two terms of (\ref{eq25h}) if we insert our result into (\ref{eq25k}).

The second integral of (\ref{eq25k}) is treated similarly and leads to the remaining two integrals of (\ref{eq25h}). 

As already mentioned in the text it is possible to derive (\ref{eq25h}) in a different way. Startingpoint is (\ref{eq25}) in which we make use of the following identity:

\begin{equation}
\int_{0}^{\infty} ds \, s^{1/2} e^{-a s} = \int_{0}^{\infty} ds \, s \frac{2}{\sqrt{\pi}} \int_{0}^{\infty} dp \, e^{-(p^{2} + a) s} = \frac{2}{\sqrt{\pi}} \int_{0}^{\infty} dp \, (p^{2} + a)^{-2}   \label{eqb4} 
\end{equation}
We get:

\begin{eqnarray}
{\cal M}^{\beta} & = & \frac{1}{\pi^{2}} m \lambda' e^{2} \epsilon_{\mu \nu \alpha \gamma} \epsilon_{1 \mu} \epsilon_{2 \nu} k_{1 \alpha} k_{2 \gamma} (2 \pi)^{4} \, \delta(\vec{k}_{1} + \vec{k}_{2} + \vec{q}) \, \delta_{\omega_{k_{1}} + \omega_{k_{2}} + \omega_{q}}  \nonumber \\
& & \int_{0}^{1} d\sigma_{1} d\sigma_{2} \int_{0}^{\infty} dp \, \left[ \frac{1}{2 \pi i} \int_{-i \infty + \mu + \epsilon}^{i \infty + \mu + \epsilon} d\omega \frac{(A + p^{2} - (\omega - i B)^{2})^{-2}}{e^{\beta (\omega -\mu)} + 1} \right.  \nonumber \\
& & \mbox{} \left. + \frac{1}{2 \pi i} \int_{-i \infty + \mu - \epsilon}^{i \infty + \mu - \epsilon} d\omega \frac{(A + p^{2} - (\omega - i B)^{2})^{-2}}{e^{- \beta (\omega - \mu)} + 1} \right] \label{eqb5}
\end{eqnarray}
where $A$ and $B$ are again given by (\ref{eq25i}). We now apply the residue theorem in order to calculate the $\omega$-integrals and use:

\begin{eqnarray}
\lefteqn{\mbox{Res}_{i B \pm \sqrt{A + p^{2}}} \left(\frac{(A + p^{2} - (\omega - i B)^{2})^{-2}}{e^{\pm \beta(\omega - \mu)}} \right)} \nonumber \\ 
& = & \mp \frac{1}{4} \left[ \frac{(A + p^{2})^{-3/2}}{e^{\pm \beta (i B \pm \sqrt{A+p^{2}} - \mu)} + 1} + \frac{\beta (A + p^{2})^{-1} e^{\pm \beta (i B \pm \sqrt{A + p^{2}} - \mu)}}{(e^{\pm \beta (i B \pm \sqrt{A + p^{2}} - \mu)} + 1)^{2}} \right]  \nonumber \\
\label{eqb6}
\end{eqnarray} 
If we rescale $p \rightarrow \frac{p}{\sqrt{A}}$ we end up with (\ref{eq25h}).

\pagebreak

\section{Some remarks about the analytical continuation}

\setcounter{equation}{0}

If we try to carry out the program proposed after formula (\ref{eq25h2}) and additionally use $k_{j}^{2} = 0$ and $(k_{j})_{0} = \frac{M}{2}$ where $M$ is the axion mass and set $M=0$ in order to get the leading order in $\frac{M}{m}$, the $\sigma$-integration becomes trivial and (\ref{eq25h}) reads (for $\mu=0$):

\begin{eqnarray}
\tilde{{\cal M}}^{\beta} & = & \frac{1}{2 \pi^{2}} m^{2} \lambda' e^{2} \epsilon_{\mu \nu \alpha \gamma} \epsilon_{1 \mu} \epsilon_{2 \nu} k_{1 \alpha} k_{2 \gamma} (2 \pi)^{4} \, \delta(\vec{k}_{1} + \vec{k}_{2} + \vec{q}) \, \delta_{\omega_{k_{1}} + \omega_{k_{2}} + \omega_{q}}  \nonumber \\
& & \int_{0}^{\infty} dp \, \left[ m^{-3} \frac{(p^{2} + 1)^{-3/2}}{1 + e^{m \beta \sqrt{p^{2} + 1}}} + \beta m^{-2} \frac{(p^{2} + 1)^{-1} e^{m \beta \sqrt{p^{2} + 1}}}{(1 + e^{m \beta \sqrt{p^{2} + 1}})^{2}} \right]   \label{eqc1}
\end{eqnarray}
where $\tilde{{\cal M}}^{\beta}$ means the leading order finite temperature contribution. We now perform a partial integration of the first term in the square bracket:

\begin{eqnarray}
\lefteqn{\int_{0}^{\infty} dp \, (p^{2} + 1)^{-1/2} \; \frac{(p^{2} + 1)^{-1}}{1 + e^{m \beta \sqrt{p^{2} + 1}}} =}  \label{eqc2} \\ 
& & \hspace{-.8cm} \int_{0}^{\infty} dp \, \mbox{arsinh}(p) \left( \frac{2 p}{(p^{2} + 1)^{2}} \frac{1}{1 + e^{m \beta \sqrt{p^{2} + 1}}} + m \beta \frac{p}{(p^{2} + 1)^{3/2}} \frac{e^{m \beta \sqrt{p^{2} + 1}}}{(1+e^{m \beta \sqrt{p^{2} + 1}})^{2}} \right)  \nonumber
\end{eqnarray}
If we use $2 \mbox{arsinh}(p)=2 \ln(\sqrt{p^{2} + 1} + p) = \ln \left( \frac{\sqrt{p^{2} + 1} + p}{\sqrt{p^{2} + 1} - p} \right)$ we arrive at:

\begin{eqnarray}
\tilde{{\cal M}}^{\beta} & = & \frac{1}{2 \pi^{2}} m^{-1} \lambda' e^{2} \epsilon_{\mu \nu \alpha \gamma} \epsilon_{1 \mu} \epsilon_{2 \nu} k_{1 \alpha} k_{2 \gamma} (2 \pi)^{4} \, \delta(\vec{k}_{1} + \vec{k}_{2} + \vec{q}) \, \delta_{\omega_{k_{1}} + \omega_{k_{2}} + \omega_{q}}  \nonumber \\
& & \int_{0}^{\infty} dp \, \left[ \frac{p}{(p^{2} + 1)^{2}} \ln \left( \frac{\sqrt{p^{2} + 1} + p}{\sqrt{p^{2} + 1} - p} \right) \frac{1}{1 + e^{m \beta \sqrt{p^{2} + 1}}}  \right. \label{eqc3} \\
& & \left. \mbox{} + m \beta \frac{1}{p^{2} + 1} \frac{e^{m \beta \sqrt{p^{2} + 1}}}{(1 + e^{m \beta \sqrt{p^{2} + 1}})^{2}} \left(1 + \ln(\sqrt{p^{2} + 1} + p) \frac{p}{\sqrt{p^{2} + 1}} \right) \right] \nonumber
\end{eqnarray}
This does not coincide with the results of [25,26]. The first term in the square bracket is exactly the known result but we get the additional second term of (\ref{eqc3}). A closer look at the calculation in the Matsubara formalism shows the following: If one continues the external Matsubara frequencies also in expressions $\exp\{-i \beta \omega_{k_{1/2}}\}$, which is indeed $1$, one obtains exactly the additional term. Such continuations play a role because the angular integration produces a $1/M^{2}$ singularity  to be cancelled by further factors $\sim \!\! M^{2}$. This subtlety in the analytical continuation should be burried in our case in the $\sigma$-integrations, which correspond to the angular integration. Unfortunately they seem to be rather difficult to handle.

\pagebreak

{\Large {\bf References}}

\vspace{1cm}

\begin{tabular}{cl}
\mbox{[1]} & R.D. Peccei and H.R. Quinn, {\em Phys. Rev. Lett.} {\bf 38} (1977), 1440; \\
& {\em Phys. Rev.} {\bf D16} (1977), 1791 \\
 & \\
\mbox{[2]} & S. Weinberg, {\em Phys. Rev. Lett.} {\bf 40} (1978), 223; \\
 & F. Wilczek, {\em Phys. Rev. Lett.} {\bf 40} (1978), 279 \\
 & \\
\mbox{[3]} & M. Dine, W. Fischler and M. Srednicki, {\em Phys. Lett.} {\bf B104} (1981), 199; \\
 & A.P. Zhitnitskii, {\em Sov. J. Nucl. Phys.} {\bf 31} (1980), 260 \\
 & \\
\mbox{[4]} & J.E. Kim, {\em Phys. Rep.} {\bf 150} (1987), 1; \\
 & M.A. Shifman, A.I. Vainshtein and V.I. Zakharov, \\
 & {\em Nucl. Phys.} {\bf B166} (1980), 493 \\
 & \\
\mbox{[5]} & R.D. Peccei, hep-ph/9606475; \\
 & R.D. Peccei in {\em CP Violation}, ed. C. Jarlskog, World Scientific \\
 & \\
\mbox{[6]} & G.G. Raffelt, {\em Phys. Rep.} {\bf 198} (1990), 1 \\
 & \\
\mbox{[7]} & M.S. Turner, {\em Phys. Rept.} {\bf 197} (1990), 67 \\
 & \\
\mbox{[8]} & G.G. Raffelt, {\em Stars as Laboratories for Fundamental Physics}, \\
 & University of Chicago Press: Chicago 1996 \\
 & \\
\mbox{[9]} & R.P. Feynman, {\em Phys. Rev.} {\bf 80} (1950), 440 \\
 & \\
\mbox{[10]} & Z. Bern and D.A. Kosower, {\em Nucl. Phys.} {\bf B379} (1992), 451 \\
 & \\
\mbox{[11]} & M.J. Strassler, {\em Nucl. Phys.} {\bf B385} (1992), 145 \\
 & \\
\mbox{[12]} & A.M. Polyakov, {\em Gauge Fields and Strings}, Harwood (1987) \\
 & \\
\mbox{[13]} & M.G. Schmidt, C. Schubert, {\em Phys. Lett.} {\bf B318} (1993), 438 \\
\end{tabular}

\begin{tabular}{cl}
\mbox{[14]} & M.G. Schmidt, C. Schubert, {\em Phys. Lett.} {\bf B331} (1994), 69; \\
 & M.G. Schmidt, C. Schubert, {\em Phys. Rev.} {\bf D53} (1996), 2150 \\
 & \\
\mbox{[15]} & M.G. Schmidt, C. Schubert, {\em Ahrenshoop Symposium} (1994), 240 \\
 & (hep-ph/9412358)\\
 & \\
\mbox{[16]} & M. Reuter, M.G. Schmidt, C. Schubert, {\em Ann. Phys.} {\bf 259} (1997), 313; \\
 & D. Fliegner, M. Reuter, M.G. Schmidt, C. Schubert, \\
 & {\em Theor. Math. Phys.} {\bf 113} (1997), 1442; \\
 & B. Kors, M.G. Schmidt, hep-th/9803144 \\
 & \\
\mbox{[17]} & D. Fliegner, P. Haberl, M.G. Schmidt, C. Schubert, hep-th/9411177 \\
 & to appear in {\em Ann. Phys.} \\
 & \\
\mbox{[18]} & S.L. Adler, C. Schubert, {\em Phys. Rev. Lett.} {\bf 77} (1996), 1695 \\
 & \\
\mbox{[19]} & M. Mondrag\'on, L. Nellen, M.G. Schmidt, C. Schubert, \\
 & {\em Phys. Lett.} {\bf B351} (1995), 200 \\
 & \\
\mbox{[20]} & M. Mondrag\'on, L. Nellen, M.G. Schmidt, C. Schubert, \\
 & {\em Phys. Lett.} {\bf B366} (1996), 212 \\
 & \\ 
\mbox{[21]} & E. D'Hoker and D.G. Gagn\'e, {\em Nucl. Phys.} {\bf B467} (1996), 272; \\
 & {\em Nucl. Phys.} {\bf B467} (1996), 297 \\
 & \\
\mbox{[22]} & N.V. Mikheev and L.A. Vassilevskaya, {\em Phys. Lett.} {\bf B410} (1997), 207 \\
 & \\
\mbox{[23]} & D.G.C. McKeon, A. Rebhan, {\em Phys. Rev.} {\bf D47} (1993), 5487 \\
 & \\
\mbox{[24]} & H. Kleinert, {\em Path Integrals}, World Scientific, chapter 6 \\
 & \\
\mbox{[25]} & S. Gupta, S.N. Nayak, hep-ph/9702205 \\
 & \\
\mbox{[26]} & A.G. Nicola and R.F. Alvarez-Estrada, {\em Z. Phys.} {\bf C60} (1993), 711 \\
 & \\
\mbox{[27]} & P. Aurenche and T. Becherrawy, {\em Nucl. Phys.} {\bf B379} (1992), 259 \\
\end{tabular}

\begin{tabular}{cl}
\mbox{[28]} & R. Kobes, {\em Phys. Rev.} {\bf D42} (1990), 562; \\
 & {\em Phys. Rev.} {\bf D43} (1991), 1269 \\
 & \\
\mbox{[29]} & T.S. Evans, {\em Phys. Lett.} {\bf B249} (1990), 286; \\
 & {\em Phys. Lett.} {\bf B252} (1990), 108 \\
 & \\ 
\mbox{[30]} & F. Guerin, {\em Nucl. Phys.} {\bf B432} (1994), 281 \\
 & \\  
\mbox{[31]} & M. Haack, {\em Rechnungen im Weltlinienformalismus bei endli-} \\
 & {\em cher Temperatur und im Magnetfeld,} \\
 & diploma thesis (with M.G. Schmidt)  \\
 & \\
\mbox{[32]} & C. Itzykson, J.-B. Zuber, {\em Quantum Field Theory}, \\
 & McGraw-Hill (1985) \\
 & \\
\mbox{[33]} & R.D. Pisarski, M.H.G. Tytgat, {\em Phys. Rev.} {\bf D56} (1997), 7077; \\
 & R.D. Pisarski, T.L. Trueman, M.H.G. Tytgat, hep-ph/9804466 \\
 & to appear in {\em Prog. Theor. Phys.} and references therein
\end{tabular}

\end{document}